\documentclass[12pt,preprint]{aastex}
\usepackage{latexsym}
\usepackage{graphicx}
\usepackage{epsfig}
\usepackage{amssymb}
\usepackage{epstopdf}

\slugcomment{Submitted for Publication in The Astrophysical Journal}
\shorttitle{Doubly Diffusive Magnetic Buoyancy}
\shortauthors{Silvers, Vasil, Brummell \& Proctor}

\begin{document}
\title{Double-diffusive instabilities of a shear-generated magnetic layer}

\author{Lara J.\ Silvers}
\affil{Department of Applied Mathematics \& Theoretical Physics, University of Cambridge, Cambridge CB3 0WA, UK}

\author{Geoffrey M.\ Vasil}
\affil{JILA, University of Colorado, Boulder, CO 80309-0440}

\author{Nicholas H.\ Brummell}
\affil{Dept.\ of Applied Mathematics \& Statistics, University of California, Santa Cruz, CA 95064}

\author{Michael R.E.\ Proctor}
\affil{Department of Applied Mathematics \& Theoretical Physics, University of Cambridge, Cambridge CB3 0WA, UK}

\begin{abstract}
Previous theoretical work has speculated about the existence of double-diffusive magnetic buoyancy instabilities of a dynamically-evolving horizontal magnetic layer generated by the interaction of forced vertically-sheared velocity and a background vertical magnetic field.  Here, we confirm numerically that if the ratio of the magnetic to thermal diffusivities is sufficiently low then such instabilities can indeed exist, even for high Richardson number shear flows.  Magnetic buoyancy may therefore occur via this mechanism for parameters that are likely to be relevant to the solar tachocline, where regular magnetic buoyancy instabilities are unlikely.
\end{abstract}


\section{Introduction}

The generation of magnetic field by velocity shear and the field's subsequent evolution are of great importance to an understanding of the operation of the solar dynamo.  While the current
dynamo paradigm contains many complex interacting components (see
e.g.\ \cite{parker_1993,rempel_2006}), an integral part of all current
large-scale solar dynamo models is the creation of strong toroidal
magnetic structures in the tachocline, the Sun's thin region of strong
radial differential rotation separating the latitudinally
differentially rotating convection zone and the solid-body-rotating radiative interior.  Strong toroidal magnetic field is thought to be induced by the stretching action of the differential rotation on any background poloidal field (the $\Omega$-effect of mean-field dynamo theory; \citet{steenbeck_etal_1966}).  Subsequently, magnetic buoyancy instabilities \citep{parker_1955} of the generated field are invoked as the mechanism for the creation of distinct magnetic structures and their subsequent rise toward eventual emergence at the solar surface as active regions.

In recent papers, \citet{vasil_brummell_2008,vasil_brummell_2009}
(hereinafter VB1, VB2) show that, when naively using the
commonly accepted paradigm for the $\Omega$-effect, it is surprisingly
difficult to initiate magnetic buoyancy instabilities if thermodynamic
adjustments are assumed to be adiabatic. Specifically, they concluded
from analytic (VB2) and numeric (VB1) calculations that for magnetic
buoyancy instabilities to operate in a thin tachocline, the velocity
shear flow imposed to drive the system must be necessarily
hydrodynamically unstable.  Roughly speaking, VB2 found that
$Ri\simeq\Delta{z}/H_{\mathrm{p}}$ is required for magnetic buoyancy,
where $Ri$ is the Richardson number, $\Delta{z}$ is the vertical shear
width and $H_{\mathrm{p}}$ is the local pressure scale height.  For
the solar tachocline, the estimated Richardson number is very large
($Ri\simeq10^{3}$--$10^{5}$) but the region is thin so that
$\Delta{z}/H_{\mathrm{p}}<1$ \citep{gough_2007} and so the above
condition for instability is therefore unlikely to be satisfied.  The
simulations of VB1 confirmed numerically that
hydrodynamically unstable imposed shear flows could generate magnetic
buoyancy instabilities in a thin shear region, but stable shear flows
did not. Energetically, this can be thought of as the following
conundrum;  for magnetic buoyancy to occur, the shear flow must
transfer enough energy into a toroidal magnetic field for it to
overcome the stable background stratification.   If the shear can only
build (through an Alfv\'{e}nic process) magnetic field to the level of
equipartition with the flow, and the shear is constrained by the stratification (since it is hydrodynamically stable), then it is difficult for the shear-induced magnetic field to overcome the constraints of the stratification.  

It is the case, however, that near the bottom of the convection zone
the ratio between the magnetic and thermal diffusivities,
$\zeta=\eta/\kappa$, is very small. It has long been recognized
\citep{gilman_1970, acheson_1979} that, in such circumstances,
instabilities can occur that rely on the much greater diffusion rate
of the stabilizing thermal component (see \cite{hughes_2007} for
further discussions). Further, VB2 noted that instability might be
enhanced by double-diffusive effects. Assuming isothermal
(as opposed to adiabatic) adjustments VB2 obtained the much less stringent
requirement that $\zeta{Ri}\simeq\Delta{z}/H_{\mathrm{p}}$. It is
therefore possible that a more-solar-like (high $Ri$) shear-generated magnetic field can become buoyantly unstable in a double-diffusive manner when $\zeta\ll1$.

In this letter, we consider this possibility in a convectively stable
 atmosphere.
 We show that if $\zeta$ is sufficiently small, and the background Alfv\'{e}nic timescales are sufficiently slow, then double-diffusive instabilities can exist for parameters that are relevant to the solar interior.

\section{Equations and parameters}

We consider a Cartesian domain $(x,y,z)$, where $z$ is depth, $x$ is
the toroidal/zonal direction, and $(y,z)$ are the poloidal
directions. Our system consists of an initially vertical
uniform magnetic field, $B_{z,0}$, permeating a stratified layer of
compressible fluid under the influence of a forcing designed to
generate a target shear flow $\mbox{\boldmath{$u$}}=(U_{0}(z),0,0)$.
This system is similar to that described in VB1 and in
\cite{silvers_etal_2009}.  We solve standard non-dimensionalized
equations for forced compressible magnetohydrodynamics governing the
evolution of velocity $\mbox{\boldmath{$u$}}=(u,v,w)$, magnetic field
$\mbox{\boldmath{$B$}}=(B_{x},B_{y},B_{z})$, density $\rho$,
temperature $T$, and pressure $P$ based around a polytropic
atmosphere,
$T_{0}(z)=1+\theta{z},\,\rho_{0}(z)=T_{0}(z)^{m},\,P_{0}(z)=T_{0}(z)^{m+1}$.
The non-dimensionalization uses $T_{*}$, the temperature at the top of
the domain, $\rho_{*}$ which is proportional to the total mass within
the domain, $P_{*}=(c_{\mathrm{p}}-c_{\mathrm{v}})T_{*}\rho_{*}$, the
fiducial pressure (given the specific heat capacities,
$c_{\mathrm{p}},c_{\mathrm{v}}$), $B_{z,0}$, the imposed background
magnetic field strength, the layer depth $d$, and time units of
$\tau_{*}=d\,\rho_{*}^{1/2}/P_{*}^{1/2}$.  We impose stress-free
velocity and vertical magnetic field boundary conditions at $z=0,1$
together with $T(z=0)=1$ and $\partial_{z}T(z=1)=\theta$. The
horizontal directions are periodic.  

The system is solved in a domain with aspect ratio 2:1:1 in a similar
manner to that in VB1 and in \cite{silvers_etal_2009} at a resolution
of (roughly) $256\times256\times512$.  This aspect ratio is specified
to allow for certain expected dynamics, e.g.\, we anticipate three-dimensional modes to
have smaller scales in the $y$ and $z$ directions than in the $x$ direction.  The thinness of the forced shear region compared to the local scale height is the geometrical factor that identifies the model with the tachocline.  The kinetic and magnetic Reynolds numbers for our calculations are roughly ${Re}\sim2000$, ${Re}_{\mathrm{M}}\sim1000$, respectively (based on the forced shear magnitude and width).  Estimates of the numerical degrees of freedom \citep{davidson_2004} imply that our resolution is reasonable.  However, we further satisfied ourselves that the results were consistent at varying resolutions.

The important parameters in the problem are the dimensionless thermal
diffusivity, $C_{K}=K\tau_{*}/\rho_{*}c_{\mathrm{p}}d^{2}$,  the
Prandtl number $\sigma=\mu{c}_{\mathrm{p}}/K$ ($\mu$ representing
dynamic viscosity), the inverse Roberts number
$\zeta=\eta{c}_{\mathrm{p}}\rho_{*}/K=\eta/\kappa$, and
$\alpha=B_{z,0}^{2}\tau_{*}^{2}/\mu_{0}\rho_{*}d^{2}$ which gives a
measure of the Alfv\'{e}n speed along the background field in terms of
the fundamental acoustic velocity scale.  Note that $\zeta$ is the critical parameter governing double-diffusive instability \citep{hughes_weiss_1995}.

We add a forcing term, $\mbox{\boldmath{$F$}}=-\sigma{C_{K}}\partial_{z}^{2}U_{0}\mbox{\boldmath{$\hat{x}$}}$, in the $x$-momentum equation that would (in the absence of magnetic effects and instabilities) maintain a desired target velocity $\mbox{\boldmath{$u$}}=(U_{0}(z),0,0)$ against viscous decay.  We choose
\begin{equation}
\label{forced shear}U_{0}(z)=\frac{1}{20}\tanh\left[10\left(z-\frac{1}{2}\right)\right]
\end{equation}
to mimic the smooth radial shear transition believed to exist in the solar tachocline. The width is chosen sufficiently narrow that
$\partial_{z}U_{0}\approx0$ (to within numerical precision) at the boundaries.

An important derived parameter is the Richardson number,
\begin{eqnarray}
Ri&=&\min_{0\le{z}\le1}\!\left(\frac{N_{0}(z)^{2}}{[\partial_{z}U_{0}(z)]^{2}}\right),
\end{eqnarray}
where $N_{0}(z)=[-g\,\partial_{z}\ln\,(P_{0}(z)^{1/\gamma}/\rho_{0}(z))]^{1/2}$
is the local Brunt-V\"{a}is\"{a}l\"{a} frequency of the stable
background atmosphere, $\partial_{z}U_{0}(z)$ is the local
turnover rate of the background shear and
$\gamma=c_{\mathrm{p}}/c_{\mathrm{v}}=5/3$.  $Ri$ measures the
relative tendency of a shear flow to overturn fluid vertically
compared to gravity's tendency to restore it to its original position.
A large $Ri$ implies that gravity is strongly stabilizing whereas a
small value can mean that shear instabilities are possible \citep{drazin_reid_2004}.  The goal in this work is to obtain a magnetic buoyancy instability at \textsl{high} Richardson number, a result not found in VB1, but here we investigate the low $\zeta$ regime where the effects of thermal stability are severely reduced.

While the Richardson number provides a useful rough measure of the stability properties of our system, we also define the respective thermal and magnetic Rayleigh numbers
\begin{eqnarray}
\label{Ra-Thermal}R_{T}(z)&=&-\frac{N_{0}(z)^{2}}{\nu(z)\kappa(z)}\\
\label{Ra-Magnetic}R_{B}(z)&=&\frac{g\alpha\,B_{x}(z)^{2}}{\nu(z)\kappa(z)P_{0}(z)}\frac{d}{dz}\ln{B_{x}(z)},
\end{eqnarray}
where $\kappa(z)=C_{K}/\rho_{0}(z)$, and $\nu(z)=\sigma\kappa(z)$.
These definitions are consistent with those given in \citet{hughes_weiss_1995} except that here the magnetic Rayleigh number, $R_{B}$, is given in a form that anticipates three-dimensional instabilities \citep{newcomb_1961}.  For the direct (non-oscillatory) type of instability, of interest here, both Rayleigh numbers are negative, and this corresponds to a thermal stratification that is stabilizing and magnetic gradient that is destabilizing.  A Rayleigh number based on the total density stratification would correspond to $R_{\mathrm{total}}=R_{T}-R_{B}$.   The key result of VB2, derived under the adiabatic thermal adjustment assumption, is equivalent to saying that $R_{\mathrm{total}}$ is always negative (ostensibly stable) unless $Ri\ll1$.  In the double-diffusive context, where thermal adjustments are closer to isothermal, the critical parameter is actually $R_{\mathrm{DD}}=R_{T}-\zeta^{-1}R_{B}$, where $\zeta\ll1$.  It is $R_{\mathrm{DD}}$ that is relevant to the stability of our system and it can potentially become positive even if $Ri\gg1$.  This is equivalent to the isothermal result derived in VB2 that required $\zeta{Ri}\simeq\Delta{z}/H_{\mathrm{p}}$.

For the basic thermal parameters (see Table~\ref{parameters}), we take
the polytropic index $m=1.6$ that enforces a convectively stable
background polytropic stratification, and a non-dimensional
lower-boundary heat flux of $\theta=5$.  With the chosen shear flow,
this produced a minimum Richardson number of $Ri=2.96$.  While this
value is significantly lower than we expect for the solar tachocline, it is high enough to guarantee that $U_{0}(z)$ hydrodynamically stable.  This fact was checked numerically with a purely hydrodynamic run.

In order to see if double-diffusive instabilities can exist, we ran two
simulations that differ \textsl{only} in the thermal diffusivity
$\kappa=K/c_{\mathrm{p}}\rho_{*}$ (varied through its non-dimensional
counterpart $C_{K}$).  Thus $\zeta$ and $\sigma$ are varied
commensurately to maintain fixed viscous and magnetic diffusivities
$\mu=\sigma{C_{K}}$ and $\eta=\zeta{C_{K}}$.  We keep
$\sigma{C_{K}}=2.5\times10^{-6}$ and $\zeta{C_{K}}=5.0\times10^{-6}$
for all our cases. For our first case, C1, anticipating instability,
we choose $\zeta=5.0\times10^{-4}$, $\sigma=2.5\times10^{-4}$ and
$C_{K}=0.01$.  In our second case, C2, anticipating stability, we choose $\zeta=0.01$, $\sigma=5.0\times10^{-3}$and $C_{K}=5.0\times 10^{-4}$.  Thus, $\zeta$ in C1 is 20-fold smaller than in C2.

The parameter  $\alpha=\sigma\zeta{C_{K}^{2}}Q$ (where $Q$ is the Chandrasekhar number measuring the ratio of magnetic and diffusive timescales) sets the background vertical magnetic field strength.  For C1 and C2 we set $\alpha=1.25\times10^{-5}$.  We desire $\alpha\ll1$ so that Alfv\'{e}n timescales are slow compared to the acoustic timescale, and $Q\gg1$ so that Alfv\'{e}n timescales are fast compared to diffusion.  C1 and C2 have $Q=1.0\times10^{6}$.  It turns out that the Alfv\'{e}n timescale of the background magnetic field has important consequences for the \textsl{dynamic} stability of our evolving MHD configuration.  To investigate this aspect, we therefore present a third case, C3, that is identical to C1 except that
$\alpha=5.0\times10^{-5}$, ($Q=4.0\times10^{6}$).  C3 therefore has two-fold faster Alfv\'{e}n timescales.

\section{Results}

\begin{figure}
\figurenum{1}
\epsscale{1}
\plotone{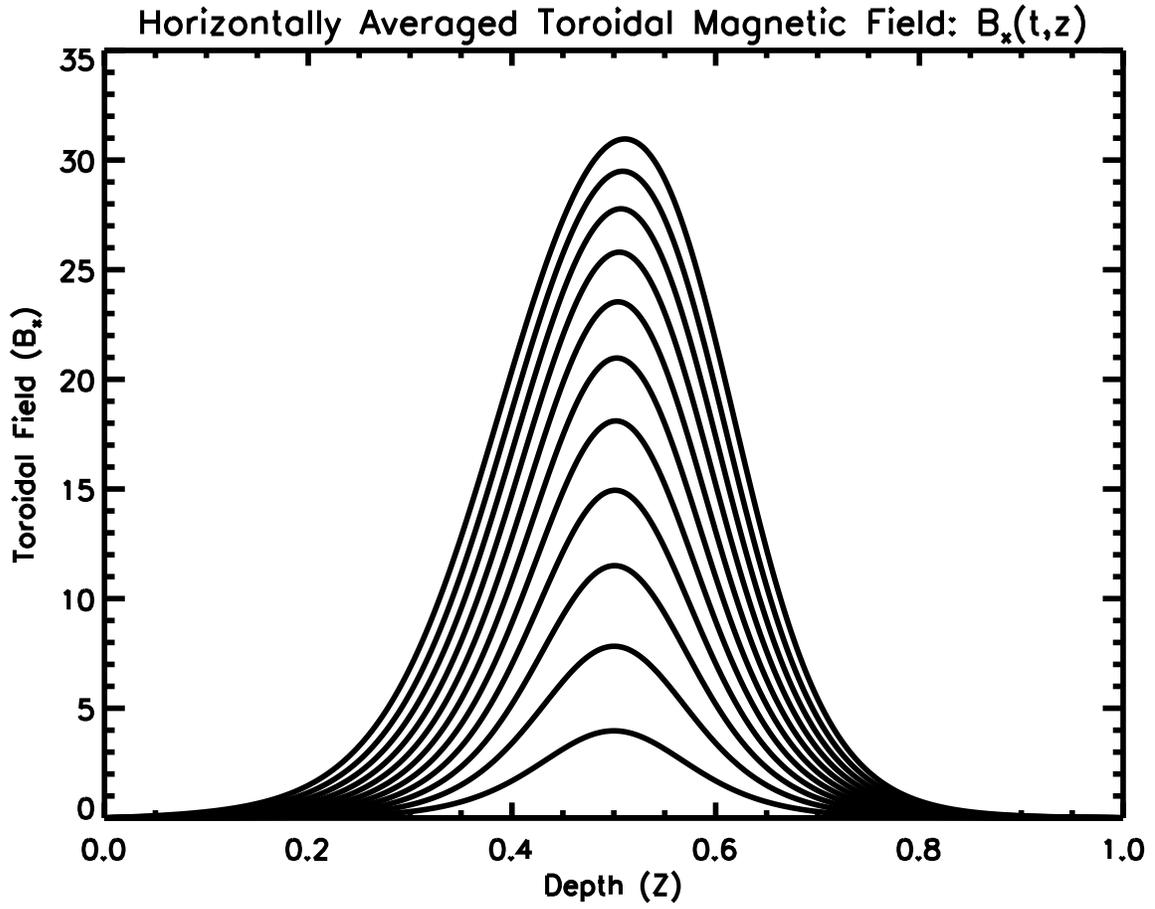}
\caption{Time evolution of growing mean toroidal magnetic field vs. depth for C1 and C2.  The field is initially zero and builds a growing peak that ends at $t\approx88$.
\label{mean-Bx}}
\end{figure}

In all three simulations, we begin with a purely vertical magnetic field, and allow the induction of a toroidal field layer by the action of the imposed velocity forcing.  In the absence of two- and three-dimensional effects, this induction process is governed by a one-dimensional (mean) set of MHD equations:
\begin{eqnarray}
\label{1D-u} \rho_{0}(z)\partial_{t}u&=&\alpha\partial_{z}B_{x}+\sigma{C_{K}}\partial_{z}^{2}u-\sigma{C_{K}}\partial_{z}^{2}U_{0}\\\label{1D-Bx}\partial_{t}B_{x}&=&\partial_{z}u+\zeta{C_{K}}\partial_{z}^{2}B_{x}.
\end{eqnarray}
Strictly speaking, the density profile in Equation~(\ref{1D-u}) will
evolve along with the shear flow and magnetic field.  However, unlike
the work in VB1-2 and \citet{silvers_etal_2009}, the target shear flow
here, $U_{0}(z)$, has a \textsl{large} Richardson number.  To an
excellent approximation, the background density maintains its initial
polytropic profile $\rho(t,z)=\rho_{0}(z)$ in this initial phase.  The
evolution, as shown in Figure~\ref{mean-Bx},  of the mean toroidal magnetic field is therefore essentially independent of the thermal properties of the system, unless  two- or three-dimensional instabilities occur. 

As time progresses, the peak field grows in the region of maximum
forced velocity shear, and the field gradients strengthen.  The
ultimate question is whether a weak shear ($Ri>1$) can induce strong
enough gradients for a magnetic buoyancy instability to occur.  Below,
we show that two initially identically evolving one-dimensional MHD configurations can have completely different stability properties, based solely on the magnitude of the thermal diffusivity.

\begin{figure*}
\figurenum{2}
\epsscale{1}
\plotone{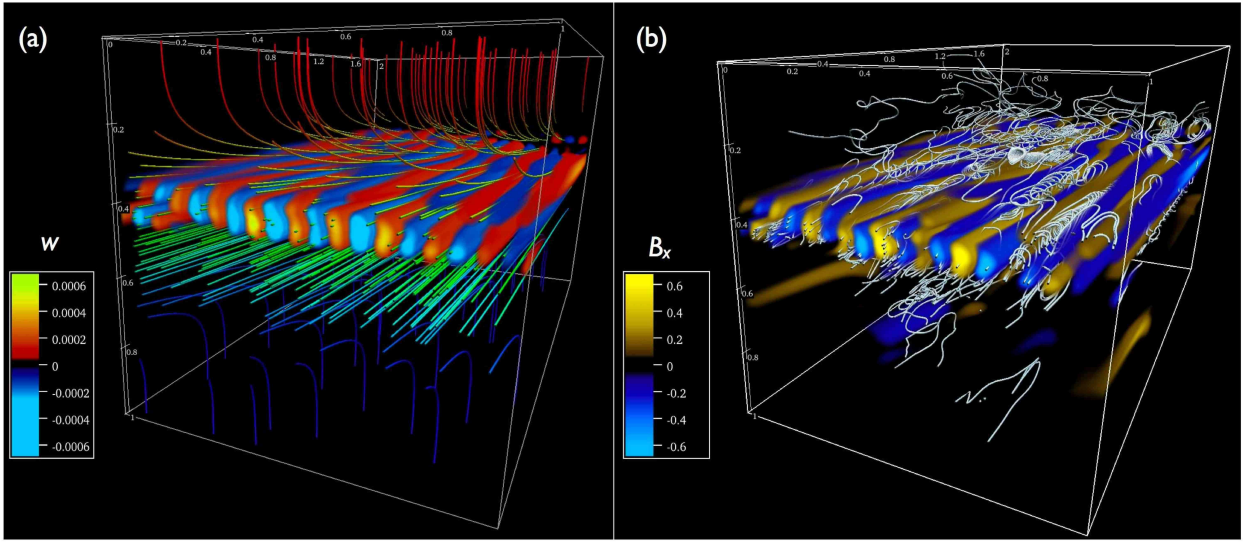}
\caption{Volume-rendered images of flow and magnetic field from C1 at $t\approx88$. Image~(a) shows vertical velocity, $w(x,y,z)$ together with lines of magnetic field colored according to zonal velocity $u(x,y,z)$ (red tones near the top of the box indicate flow of one direction, blue tones near the bottom of the box indicate opposite flow, and green tones indicate approximate stagnation).   Image~(b) shows fluctuating toroidal magnetic field, $B_{x}(x,y,z)-\overline{B}_{x}(z)$.  The silver lines indicate streamlines of the fluctuating velocity $(u-U_{0},v,w)$.
\label{volumes}}
\end{figure*}

\begin{figure}
\figurenum{3}
\epsscale{1}
\plotone{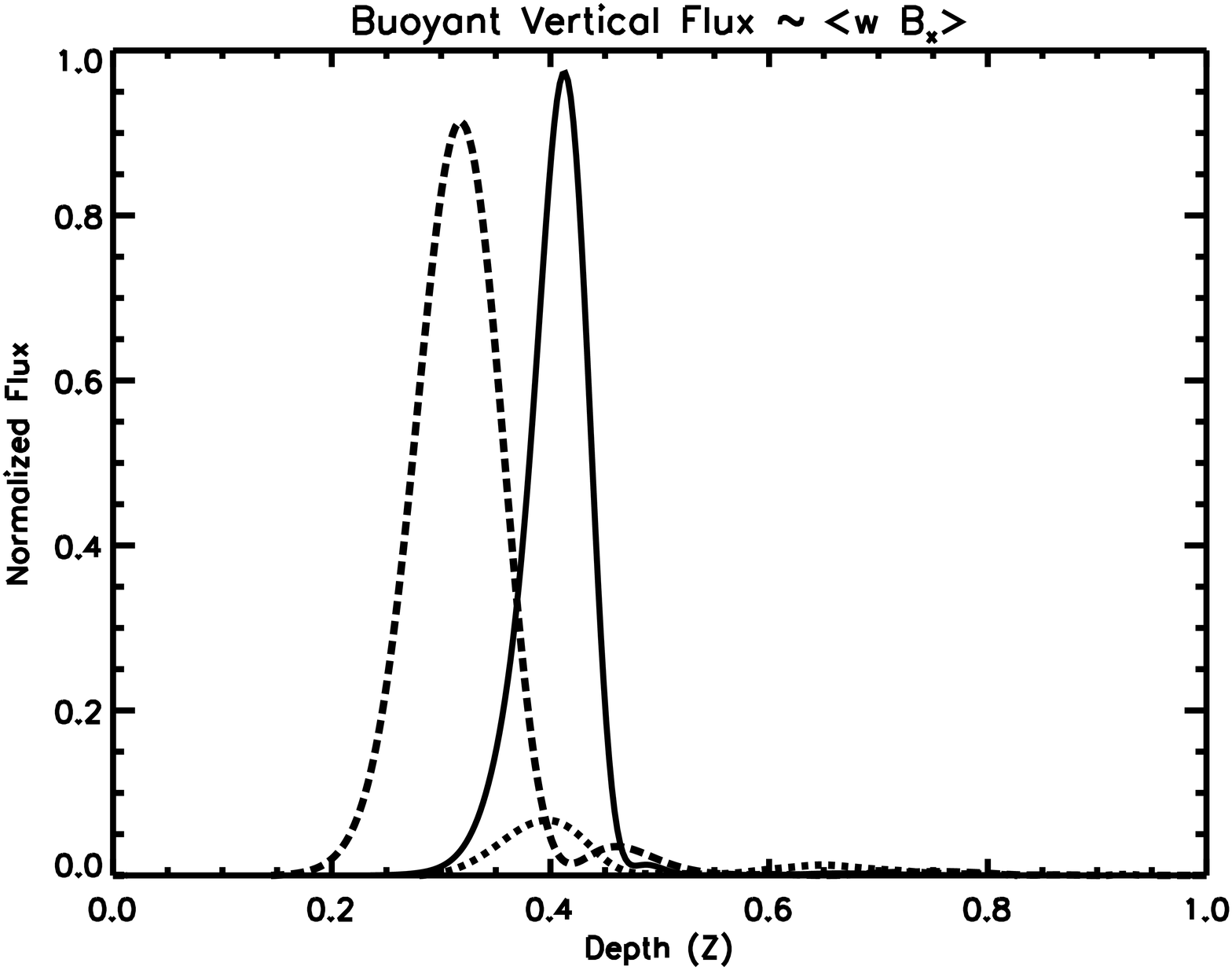}
\caption{Normalized buoyant vertical flux vs. depth for C1 (solid) and C3 (dotted or dashed).  All plots are proportional to $\overline{w{B_{x}}}-\overline{w}\overline{B}_{x}$ for each simulation and are scaled to be in the same units ($B_x$ rescaled, and all normalized by the same constant value ($\max_{z}\left[\left(\overline{w^{2}}-\overline{w}^{2}\right)\left(\overline{B_{x}^{2}}-\overline{B}_{x}^{2}\right) \right]^{1/2}$ from C1).  Since the Alfv\'{e}n timescales are twice as fast, the C3 plots are shown both when the instability is occurring in the same vertical location as C1 (dotted, $t\approx44$) and at the same actual time as the C1 plot (dashed, $t\approx88$).  
\label{flux-correlation}}
\end{figure}

In C1, the system does indeed initially evolve according to
Equations~(\ref{1D-u})~and~(\ref{1D-Bx}).  However, since
$\zeta=5.0\times10^{-4}$ is small enough, the induction of mean
toroidal field only proceeds for a finite time before an instability
becomes apparent at $t\approx88$.  Figure~(\ref{volumes}) shows
volume-rendered images at this time of the vertical velocity
$w(x,y,z)$ and the fluctuating toroidal magnetic field
$B_{x}(x,y,z)-\overline{B}_{x}(z)$.  In the initial induction phase
(governed by Equations~(\ref{1D-u})~and~(\ref{1D-Bx})) both these
quantities are zero. However, Figure~(\ref{volumes}) clearly shows
that a wave-like perturbation in the vertical velocity and toroidal
field has appeared in the region of strong toroidal magnetic field
gradients near the top of the evolving toroidal magnetic layer.  The
instability appears in a quasi-two-dimensional manner with a wavevector primarily
in the $y-z$ plane, taking a form that is similar to a classical
`interchange' instability \citep{cattaneo_hughes_1988}.   Roll-like
motions principally swap lines of toroidal magnetic field that pierce
$y-z$ planes without a large degree of bending along the toroidal
$x$-direction.  There is a high degree of correlation between vertical
velocity and toroidal field perturbations, strongly implying a
buoyancy-driven instability (also see Figure~\ref{flux-correlation}).

We now compare C2 results with those of C1, which differs only the value of
$\zeta$.  We estimate the value of $\zeta$ for C2 to be sufficiently
large to render $R_{\mathrm{DD}}$ everywhere negative, and therefore
anticipate that the instability found in C1, if double-diffusive,
should not appear in this case.  We run C2 for over twice the time that it takes for the instability in C1 to manifest (up to $t\approx180$) and the system simply continues to evolve according to the mean Equations~(\ref{1D-u})~and~(\ref{1D-Bx}).  Eventually, these dynamics are benignly disrupted by Alfv\'{e}nic processes without any buoyancy instabilities (as explained in VB2) and we therefore halt the computation.    We conclude that the instability discovered in C1 is of a double-diffusive nature, since it depends critically on a sufficiently large thermal diffusivity (small $\zeta$).

A subtle issue in this problem is the influence of the strength of the imposed background field.  As argued in VB1-2, this essentially provides a timescale for disruptive
Alfv\'{e}nic processes.   For instability to occur, the necessary
conditions must be met before Alfv\'{e}nic processes can disrupt the
source.  In VB1-2, this required a {\sl strong} velocity shear
(hydrodynamically unstable, $Ri\approx0.03$) in order to induce the
required magnetic gradients sufficiently quickly for a regular
(mean-density-deficit driven) magnetic buoyancy instability to
occur. Things are more complex in the high $Ri$, double-diffusive regime considered
here. To elucidate this issue, we examine case C3 where the imposed background field strength is twice that of C1 and C2.   Figure~(\ref{flux-correlation}) shows a vertical profile of the mean vertical transport of $B_x$ for both C1 and C3, a good indicator of the existence and efficiency of any magnetic buoyancy instabilities.  The measure plotted is
\begin{eqnarray}
C(z)&=&\frac{\overline{w{B_{x}}}-\overline{w}\overline{B}_{x}}{\max_{z}|w_{\mathrm{rms}}B_{x,\mathrm{rms}}|},
\end{eqnarray}
where the overline represents a horizontal average, and $w_{\mathrm{rms}}(z)=\overline{w^{2}}-\overline{w}^{2}$, $B_{x,\mathrm{rms}}(z)=\overline{B_{x}^{2}}-\overline{B}_{x}^{2}$.  
For ease of comparisons, the normalization factor in the denominator for all cases is that value
calculated for C1 and $B_{x}$ is rescaled to the same units.  Figure~(\ref{flux-correlation}) shows
that the toroidal magnetic field perturbations correlate well with
vertical velocity perturbations, confirming that the instabilities are
most likely of a magnetic buoyancy type.   However, the dotted curve in this figure shows that the transport is significantly weaker for C3 in the initial stages of the instability. 
This is a signature of the dynamic nature of the instability.  Owing
to the evolving background state, $R_{\mathrm{DD}}$ does not entirely
determine stability.  In the
equilibrium stability problem for quasi-incompressible motion without
an imposed shear considered in \citet{hughes_weiss_1995},
the growth rate of the instability becomes positive when
$R_{\mathrm{DD}}\ge27\pi^{4}/4$.  In the current problem, the portion
of the domain that is unstable is perpetually moving as the
magnetic layer evolves.  Therefore, a locally unstable
perturbation may not have enough time to amplify to a dynamically
significant magnitude before it finds itself outside of the moving
region of instability.  Only if the growth rate of the instability is
sufficiently large can an unstable mode amplify fast enough that the background appears effectively steady.  Therefore, to exhibit instability in this evolving system unambiguously, the growth rate of the instability must be at least of the order of the Alfv\'{e}nic evolution timescale for the background,
\begin{eqnarray}
\label{growth-rate}p_{\mathrm{crit}}(R_{T},R_{B},\sigma,\zeta)\approx\Delta{z}\sqrt{\sigma\zeta{Q}},
\end{eqnarray}
where, $p_{\mathrm{crit}}(R_{T},R_{B},\sigma,\zeta)$ is the maximal
growth rate as a function of the Rayleigh numbers as given in
\citet{hughes_weiss_1995}, and the Chandrasekhar number $Q\gg1$.
Equation~(\ref{growth-rate}) implies that in the presence of a
stronger background field, the instability requires a faster growth
rate to proceed in the same manner.  Even
though C3 has the exact same background $\overline{B}_{x}(z)$
configuration as that for C1, the stronger background field strength,
$B_{z,0}$, alters the nature of the instability.   At the location
where C1 grows strongly, C3 only reaches small amplitude (albeit in a
shorter time: $t\approx44$ \textit{cf}.\ $t\approx88$).  However, the
  instability in C3 can still grow to significant amplitudes {\sl
    higher up} in the domain as the system evolves further, as shown
  by the dashed curve in Figure~(\ref{flux-correlation}).   The
  amplitude is further modified since the stable background
  density stratification, and therefore $R_{\mathrm{DD}}$ changes slightly with height.  

\begin{figure}
\figurenum{4}
\epsscale{1}
\plotone{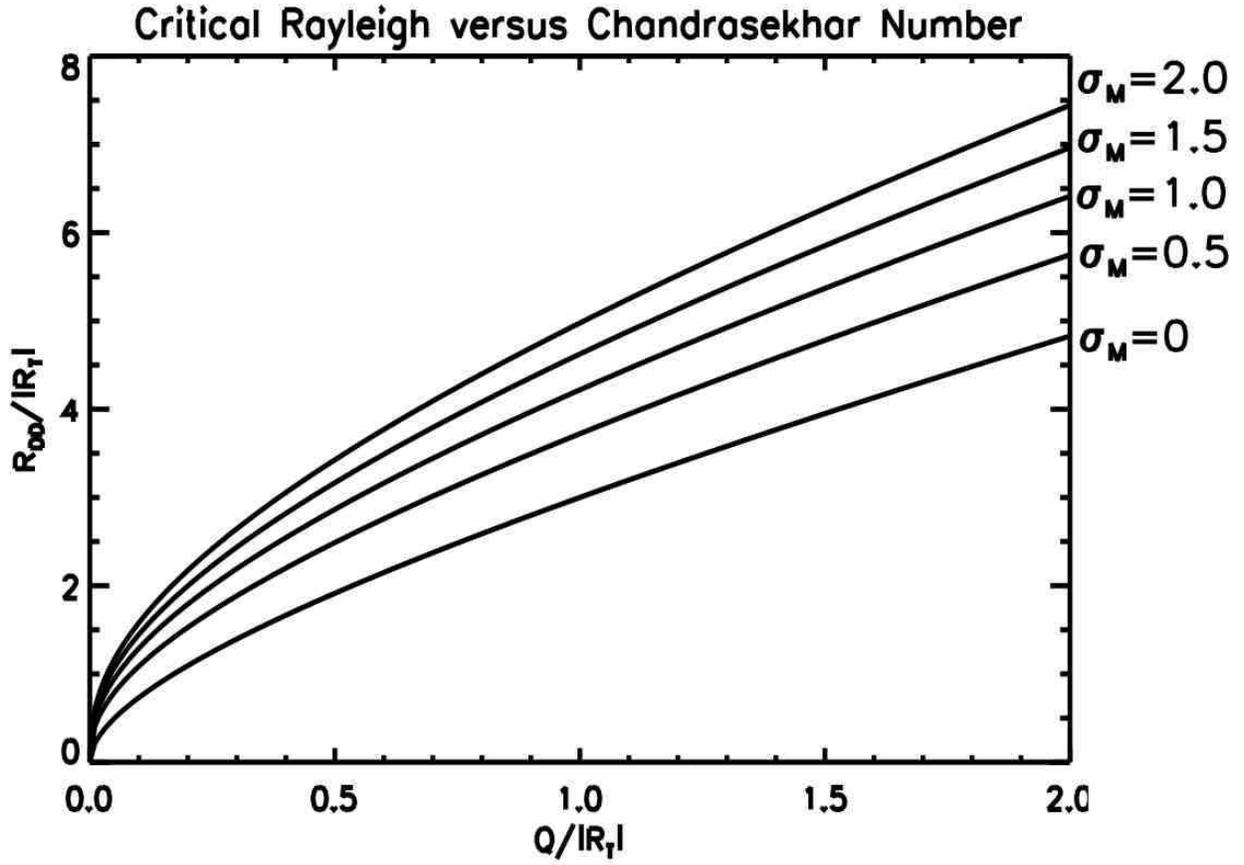}
\caption{Critical $R_{DD}/|R_{T}|$ vs. $Q/|R_{T}|$ for $\sigma_{\mathrm{M}}=\sigma/\zeta=0,0.5,1.0,1.5,2.0$, in the limits $\sigma,\zeta\ll1$, and $R_{T}\ll0$.
\label{critical-parameter}}
\end{figure}

We have demonstrated that the presence of shear-generated double-diffusive magnetic buoyancy instabilities can be controlled primarily by the single parameter $\zeta$.  Given the dynamic nature of the instability here, the second most important parameter is the Chandresekhar number, $Q$, which dictates the background Alfv\'{e}nic timescales.  C3 provides good evidence that the efficiency of the instability is governed through a relation at least qualitatively similar to Equation~(\ref{growth-rate}).  However, the question remains as to whether any other parameters, such as the magnetic Prandtl number, $\sigma_{\mathrm{M}}=\sigma/\zeta$, are critical to this process.   Numerical constraints restrict our simulations to $\sigma_{\mathrm{M}}=0.5$, which is less than unity thus preserving the correct ordering of timescales, but larger than the estimated solar value of $\sigma_{\mathrm{M}}\approx6\times10^{-2}$ \citep{gough_2007}.
To make some progress, we extend the analysis of \citet{hughes_weiss_1995}, noting again that this discusses the double-diffusive 
instabilites of an equilibrium rather than a dynamically-evolving system.  It can be shown that Equation~(\ref{growth-rate}) embodies the solution to a $12^{\mathrm{th}}$-order polynomial in $p_{\mathrm{crit}}$.   In the limit $\sigma,\zeta\ll1$ and $R_{T}\ll0$ this can be simplified to a cubic polynomial depending solely on the reduced variables $r = R_{\mathrm{DD}}/|R_{T}|$, $q=\Delta{z}^{2}Q/|R_{T}|$, and $\sigma_{\mathrm{M}}$:
\begin{eqnarray}
&&4X_{r}^{3}+X_{q}X_{r}^{2}-18X_{q}X_{r}-X_{q}(27+4X_{q})=0,
\end{eqnarray}
where
\begin{eqnarray}
X_{q}=\frac{q(1+\sigma_{\mathrm{M}})^{3}}{\sigma_{\mathrm{M}}^{2}},\quad{X}_{r}=\frac{(r-q)(1+\sigma_{\mathrm{M}})}{\sigma_{\mathrm{M}}}.
\end{eqnarray}
Figure~(\ref{critical-parameter}) shows the resulting dependence of
the critical value of $r$ versus $q$ for several values of
$\sigma_{\mathrm{M}}$.  In particular, it can be seen that the
instability is more easily obtained for smaller $\sigma_{\mathrm{M}}$,
which makes sense in terms of the lessening relative importance of
viscosity, and, encouragingly, supports the viability of this process
at more realistic solar parameters.

The transport measure
$\overline{w{B_{x}}}-\overline{w}\overline{B}_{x}$ has a more
complicated dependence on $\sigma_{\mathrm{M}}$.  In the appropriate
Rayleigh number regime,
$\overline{w{B_{x}}}-\overline{w}\overline{B}_{x}\sim{p_{\mathrm{crit}}}+k_{\mathrm{crit}}^{2}$,
where $k_{\mathrm{crit}}$ is the horizontal wavenumber of the fastest
growing mode.  As $\sigma_{\mathrm{M}}$ is decreased,
$k_{\mathrm{crit}}$ decreases and $p_{\mathrm{crit}}$ increases, and
hence the dependence on $\sigma_{\mathrm{M}}$ is not straightforward.
However, in the limit as $\sigma_{\mathrm{M}}\rightarrow0$, then
$\overline{w{B_{x}}}-\overline{w}\overline{B}_{x}\sim{p_{\mathrm{crit}}}$,
which, from Figure~(\ref{critical-parameter}), is nonzero even at
$\sigma_{\mathrm{M}}=0$.  This argument implies that, if we were able
to run a computation at a significantly lower value of
$\sigma_{\mathrm{M}}$ than in C1 and C3 (an extreme numerical
challenge given our other parameters), we might expect that the buoyant
transport should decrease, but not disappear altogether.  

\section{Conclusions}

After the pessimistic results of VB1-2, it is encouraging that when
double-diffusive effects are taken into account, the dynamics
intuitively expected in the tachocline appear to be possible,
satisfying the constraints that we have anticipated.  We know roughly
that $\zeta\approx10^{-5}$ and $Ri\approx10^{3}\!-\!10^{5}$ in the tachocline, and it is therefore plausible that their product is smaller than $\Delta{z}/{H}_{\mathrm{p}}\approx1$ \citep{gough_2007}.  However, we know very little about the configuration of any magnetic field in the solar tachocline or the associated Alfv\'{e}nic timescales, and hence we do not know exactly {\sl how much} smaller $\zeta$ must be for the growth rate of an instability to become large enough for it to proceed.

Even if instability occurs, it still remains unclear whether the nonlinear evolution of the system can produce magnetic fields that are sufficiently strong and coherent to be able to rise through the turbulent convection zone.  
While it now seems possible that high $Ri$ shear flows could produce
buoyant magnetic structures in the tachocline, one might expect that 
such buoyant structures would have magnetic energies that are (at
most) in equipartition with the shear.  If the turbulent convection zone acts locally in maintaining the tachocline's shear, then it is difficult to see how magnetic structures could obtain the significantly greater strengths that are required to survive the disrupting effects of the convection zone \citep{tobias_etal_2001,cline_2003,fan_etal_2003}.  Of course, this question depends critically on how the energy of the shear relates to that of the convection zone, which is not well understood.

\clearpage

\begin{deluxetable}{ccccccc}
 \tabletypesize{\footnotesize}
  \tablecolumns{6}
  \tablewidth{0pt}
  \tablecaption{Parameters for Simulations
  \label{parameters}}
  \tablehead{\colhead{Case}  &    \colhead{ $\zeta$}
                                              &    \colhead{$\sigma$}
                                              &    \colhead{$C_{K}$}
                                              &    \colhead{$\alpha$}
                                              &    \colhead{$Q$}
                                              &    \colhead{Stability}
 }
 \startdata
               C1 &
                $5.0 \times 10^{-4}$&
                $2.5 \times 10^{-4}$&
                $1.0 \times 10^{-2}$&
                $1.25 \times 10^{-5}$&
                $1.0 \times 10^{6}$&
                 unstable\\
               C2 &
                $1.0 \times 10^{-2}$&
                $5.0 \times 10^{-3}$&
                $5.0 \times 10^{-4}$&
                $1.25 \times 10^{-5}$&
                $1.0 \times 10^{6}$&     
                 stable\\
               C3 &
                $5.0 \times 10^{-4}$&
                $2.5 \times 10^{-4}$&
                $1.0 \times 10^{-2}$&
                $5.0 \times 10^{-5}$&
                $4.0 \times 10^{6}$&
                 delayed onset
   \enddata
   \tablecomments{In all cases, $m=1.6$, $\theta=5$, and $Ri_{\min}=2.96$}
\end{deluxetable}

\clearpage

\clearpage

\clearpage

\end{document}